\documentclass[traditabstract]{aa}
\bibpunct{(}{)}{;}{a}{}{,} 
\usepackage{graphicx}
\usepackage[varg]{txfonts}
\usepackage{color}

\begin{document}

\title{Polarization of the changing-look quasar J1011$+$5442\thanks{Based on observations made with the William Herschel telescope operated on the island of La Palma by the Isaac Newton Group of Telescopes in the Spanish Observatorio del Roque de los Muchachos of the Instituto de Astrofísica de Canarias.}}
\author{D. Hutsem\'ekers\inst{1,}\thanks{Senior Research Associate F.R.S.-FNRS},
        B. Ag{\'\i}s Gonz\'alez\inst{1,2},
        D. Sluse\inst{1},
        C. Ramos Almeida\inst{4,5},
        J.-A. Acosta Pulido\inst{4,5}
        }
\institute{
    Institut d'Astrophysique et de G\'eophysique,
    Universit\'e de Li\`ege, All\'ee du 6 Ao\^ut 19c, B5c,
    4000 Li\`ege, Belgium
    \and 
    Instituut voor Sterrenkunde, KU Leuven, Celestijnenlaan 200D, Bus 2401, 3001, Leuven, Belgium
    \and
    Instituto de Astrofisica de Canarias, Calle Via Lactea, s/n, E-38205 La Laguna, Tenerife, Spain
    \and
    Departamento de Astrofisica, Universidad de La Laguna, E-38205 La Laguna, Tenerife, Spain
    }
\date{Received ; accepted: }
\titlerunning{Polarization of changing-look quasars} 
\authorrunning{D. Hutsem\'ekers et al.}
\abstract{If the disappearance of the broad emission lines observed in changing-look quasars were caused by the obscuration of the quasar core through moving dust clouds in the torus, high linear polarization typical of type~2 quasars would be expected. We measured the polarization of the changing-look quasar J1011$+$5442 in which the broad emission lines have disappeared between 2003 and 2015.  We found a polarization degree compatible with null polarization. This measurement suggests that the observed change of look is not due to a change of obscuration hiding the continuum source and the broad line region, and that the quasar is seen close to the system axis. Our results thus support the idea that the vanishing of the broad emission lines in J1011$+$5442 is due to an intrinsic dimming of the ionizing continuum source that is most likely caused by a rapid decrease in the rate of accretion onto the supermassive black hole.
}
\maketitle
%
%
%
\section{Introduction}
\label{sec:intro}
Type~1 quasars and lower luminosity active galactic nuclei (AGN) are characterized by broad emission lines (BELs) in their optical spectrum.  These lines reveal high-velocity motions in the vicinity of the accretion disk that feeds the central supermassive black hole and emits the continuum spectrum. The high luminosity and strong outflows generated by quasar activity can be powerful enough to regulate star formation in the host galaxy and then play a significant role in their cosmological evolution \citep{1998Silk,2013Borguet}.  The timescale of the activity cycle(s) in galaxies is thus a key parameter. Indirect measurements suggest that the quasar phenomenon is a relatively short stage in the evolution of galaxies, of the order of 10$^{7-8}$ years \citep{2001Martini}. \citet{2015Schawinski} estimated AGN lifetimes of the order of 10$^{5}$ years, implying that black holes probably grow through short bursts. Quasars are then expected to ``flicker'' on and off.

Although transition objects are expected to be elusive, a handful of them may have been discovered recently. \citet{2015LaMassa} found the first quasar changing from type~1 (strong broad and narrow emission lines) to type~1.9 (strong narrow lines only and dim continuum). Soon after this, \citet{2016Runnoe} identified another quasar in which the BELs had disappeared.  Thanks to systematic searches in the spectroscopic archive of the Sloan Digital Sky Survey (SDSS), \citet{2016Ruan} and \citet{2016MacLeod} uncovered about ten new changing-look quasars. The timescales of the changes are of the order of only a few years.

To explain these spectral changes, two main interpretations have been proposed.  (I) Modifications in the source of ionizing radiation, likely a variation in accretion rate onto the supermassive black hole \citep{2014Elitzur}. An intrinsic dimming of the continuum source reduces the number of photons available to ionize the gas, resulting in a net decrease in emission line intensity. (II) Variable dust absorption along the observer's line of sight to the continuum source and the broad line region that is due to the motion of individual dust clouds in a clumpy torus \citep{1989Goodrich,1992Tran}.

These two possible scenarios can be better distinguished in high-luminosity AGNs (quasars) because the inner BEL region is larger in this class, so that the lower limit on the crossing time of dust clouds consequently increases \citep{2015LaMassa}. The size of the broad line region is indeed proportional to $L^{\alpha}$ with $\alpha = 0.5-0.7$, $L$ being the 5100 \AA\ optical luminosity \citep{2005Kaspi,2006Bentz}. It follows that the variable dust absorption scenario seems to be disfavoured in quasars because the extinction variation timescales that are due to dust clouds moving in front of the broad line region are factors 2-10 too long to explain the observed spectral changes \citep{2015LaMassa,2016MacLeod}. These results make changing-look quasars good candidates to be flickering AGN, and thus possible Rosetta stones for the study of quasar activity cycles.

In the present paper we use the low- / high-polarization dichotomy between type~1 and~2 quasars to further constrain these scenarios. In Sect.~\ref{sec:pola} we specify the polarization dichotomy between type~1 and~2 quasars in the redshift and luminosity ranges of interest. In Sect.~\ref{sec:obs} we report the polarization measurement of the changing-look quasar J1011$+$5442, in which broad emission lines have recently disappeared. We conclude in Sect.~\ref{sec:end} that the null-polarization measured in J1011$+$5442 does not support the variable dust absorption scenario.

\section{Polarization of type~1 and~2 quasars}
\label{sec:pola}

\begin{table}[t]
\caption{Polarization of type~2 and~1 quasars}
\label{tab:pola}
\centering
\begin{tabular}{lclrrr}
\hline\hline
Object & $z$ & Type & $p$~~ & $\sigma_p$ & Ref. \\
\hline
J081507.42$+$430427.2 &  0.510 &  ~~~2  &    8.5 &   0.6  &   Z05 \\
J084234.94$+$362503.1 &  0.561 &  ~~~2\tablefootmark{*} &   16.5 &   0.2  &   Z05 \\
J092014.11$+$453157.2 &  0.402 &  ~~~2  &    4.7 &   0.9  &   Z06 \\
J103951.49$+$643004.2 &  0.402 &  ~~~2\tablefootmark{*} &   16.6 &   0.3  &   Z05 \\ 
J110621.95$+$035747.1 &  0.242 &  ~~~2  &    3.1 &   0.5  &   Z06 \\  
J132323.33$-$015941.9 &  0.350 &  ~~~2  &    5.0 &   1.0  &   Z05 \\ 
J140740.06$+$021748.3 &  0.309 &  ~~~2  &    2.6 &   0.2  &   Z05 \\ 
J141315.31$-$014221.0 &  0.380 &  ~~~2  &    4.1 &   1.0  &   Z05 \\  
J150608.09$-$020744.2 &  0.439 &  ~~~2  &    9.7 &   1.1  &   Z05 \\  
J151711.47$+$033100.2 &  0.613 &  ~~~2  &   10.5 &   0.8  &   Z05 \\ 
J154340.02$+$493512.6 &  0.512 &  ~~~2\tablefootmark{*} &    7.7 &   0.3  &   Z05 \\  
J164131.73$+$385840.9 &  0.596 &  ~~~2\tablefootmark{*} &    4.5 &   0.3  &   Z05 \\ 
J171559.79$+$280716.8 &  0.524 &  ~~~2\tablefootmark{*} &    3.5 &   0.3  &   Z05 \\
\hline
J014017.06$-$005003.0 &  0.334 & ~~~1\tablefootmark{$\dagger$}  &   0.63 &   0.31 &   S84 \\ 
VJ015130.8$+$091725   &  0.299 & ~~~1  &   1.21 &   0.54 &   M84 \\ 
J095048.38$+$392650.4 &  0.205 & ~~~1\tablefootmark{$\dagger$}  &   0.19 &   0.15 &   B90 \\ 
J095652.39$+$411522.2 &  0.234 & ~~~1  &   0.25 &   0.22 &   B90 \\
J105151.44$-$005117.6 &  0.359 & ~~~1\tablefootmark{$\dagger$}  &   0.57 &   0.28 &   B90 \\ 
J112439.18$+$420145.0 &  0.225 & ~~~1\tablefootmark{$\dagger$}  &   0.44 &   0.19 &   B90 \\
J113230.09$-$024620.9 &  0.332 & ~~~1  &   0.53 &   0.16 &   S05 \\ 
J113432.29$-$005548.1 &  0.268 & ~~~1\tablefootmark{$\dagger$}  &   0.42 &   0.18 &   S05 \\ 
J121920.93$+$063838.5 &  0.331 & ~~~1  &   0.60 &   0.13 &   S05 \\ 
J121946.53$+$145259.3 &  0.401 & ~~~1\tablefootmark{$\dagger$}  &   0.52 &   0.26 &   S05 \\ 
J124654.03$+$131310.7 &  0.512 & ~~~1\tablefootmark{$\dagger$}  &   0.14 &   0.14 &   S05 \\ 
J124730.96$-$014227.7 &  0.346 & ~~~1\tablefootmark{$\dagger$}  &   1.12 &   0.36 &   S05 \\ 
J130112.91$+$590206.6 &  0.477 & ~~~1  &   0.06 &   0.17 &   B90 \\ 
J134251.60$-$005345.3 &  0.326 & ~~~1\tablefootmark{$\dagger$}  &   0.21 &   0.14 &   S05 \\ 
J135632.79$+$210352.3 &  0.300 & ~~~1\tablefootmark{$\dagger$}  &   1.42 &   0.31 &   B90 \\
J140031.96$+$040457.6 &  0.428 & ~~~1\tablefootmark{$\dagger$}  &   0.47 &   0.27 &   S84 \\ 
J142943.07$+$474726.2 &  0.220 & ~~~1\tablefootmark{$\dagger$}  &   0.25 &   0.29 &   B90 \\ 
J144645.93$+$403505.7 &  0.267 & ~~~1  &   0.37 &   0.15 &   B90 \\ 
J154530.23$+$484609.0 &  0.401 & ~~~1  &   0.08 &   0.20 &   B90 \\ 
J161410.62$+$263250.4 &  0.395 & ~~~1  &   1.24 &   0.56 &   S84 \\ 
VJ214423.0$+$041627   &  0.463 & ~~~1  &   0.84 &   0.25 &   S05 \\ 
VJ215819.1$+$020849   &  0.560 & ~~~1  &   0.51 &   0.22 &   S05 \\ 
VJ222049.7$-$315654   &  0.506 & ~~~1  &   0.55 &   0.28 &   S05 \\
VJ232210.8$-$344757   &  0.420 & ~~~1  &   0.32 &   0.22 &   S05 \\ 
VJ234844.8$-$361827   &  0.541 & ~~~1  &   0.64 &   0.25 &   S05 \\ 
VJ235729.3$+$002940   &  0.410 & ~~~1  &   0.67 &   0.30 &   S05 \\ 
\hline
\end{tabular}
\tablefoot{Object name and coordinates are either from SDSS or from \citet{2001Veron}, the latter with prefix V. The linear polarization degree $p$ and its error $\sigma_p$ are given in \%.
\tablefoottext{*}{Type~2 quasars for which broad emission lines have been detected in the polarized light.}
\tablefoottext{$\dagger$}{Type~1 quasars with L$_{\text{[\ion{O}{iii}]}}$ $\gtrsim$ 3$\times$10$^8$ L$_{\sun}$.}}
\tablebib{(Z05)~\citet{2005Zakamska}; (Z06) \citet{2006Zakamska}; (S84) \citet{1984Stockman}; (M84) \citet{1984Moore}; (B90) \citet{1990Berriman}; (S05) \citet{2005Sluse}.}
\end{table}

Quasar light is known to be linearly polarized at optical wavelengths, with significant differences between type~1 and~2 objects, the latter showing the highest polarizations.

\citet{2003Zakamska} have built a large sample of type~2 quasars from the spectroscopic data of the Sloan Digital Sky Survey. These objects are optically faint, with no broad permitted lines, and show narrow emission lines with high equivalent widths. Since the broad line region and the source of continuum are obscured, the luminosity of the [\ion{O}{iii}] lines, assumed to come from an extended and therefore unobscured region, is used as a proxy of the intrinsic luminosity. Since luminous (M$_{\text B} <$ $-$23) type~1 quasars have, on average,  [\ion{O}{iii}] $\lambda$5007 line luminosities higher than 3$\times$10$^8$ L$_{\sun}$, this value was used as a cutoff to identify intrinsically luminous type~2 objects \citep{2003Zakamska}. Subsequent spectropolarimetry \citep{2005Zakamska,2006Zakamska} has revealed highly polarized quasars with broad lines in their polarized spectrum, which clearly proved that these type~2 quasars indeed harbour a luminous quasar in their highly obscured core. In Table~\ref{tab:pola} we have collected the linear polarization degree ($p$ and its error $\sigma_p$ in \%) of the radio-quiet type~2 quasars studied in \citet{2005Zakamska,2006Zakamska}. Only the rest-frame UV-blue (2820-3710 \AA ) polarization  measurements are considered here. All these measurements have $p/\sigma_p > 4$. The sample contains 13 type~2 quasars with redshifts $z$ between 0.2 and 0.7.

To build a comparison sample of type~1 quasar polarizations, we used the data set compiled by \citet{2005Hutsemekers}. We restricted the sample to good-quality polarization data, that is, we only considered objects with either $p \geq 0.6\%$ and $p/\sigma_p \geq 2$, or  $p < 0.6\%$ and $\sigma_p < 0.3\%$. To minimize contamination by interstellar polarization, we also selected objects at high Galactic latitudes $|b_{\text{gal}}| \geq 30\degr$. Radio-loud and red quasars were discarded, as were broad absorption line quasars, the polarization of which differs from that of normal radio-quiet objects \citep{1984Moore,1998bHutsemekers,2002Smith}. The final sample, matching the redshift range of type~2 quasars $0.2 < z < 0.7$, contains 26 radio-quiet type~1 quasars (Table~\ref{tab:pola}). All these objects are type~1 quasars with M$_{\text B} <$ $-$23 \citep{2001Veron}. Fourteen of them have measured [\ion{O}{iii}] $\lambda$5007  luminosities that are between 1$\times$10$^8$ and 8$\times$10$^9$ L$_{\sun}$ \citep{2011Shen}, 12 of them satisfying L$_{\text{[\ion{O}{iii}]}}$ $\gtrsim$ 3$\times$10$^8$ L$_{\sun}$ within the uncertainties.

\begin{figure}[t]
\centering
\resizebox{\hsize}{!}{\includegraphics*{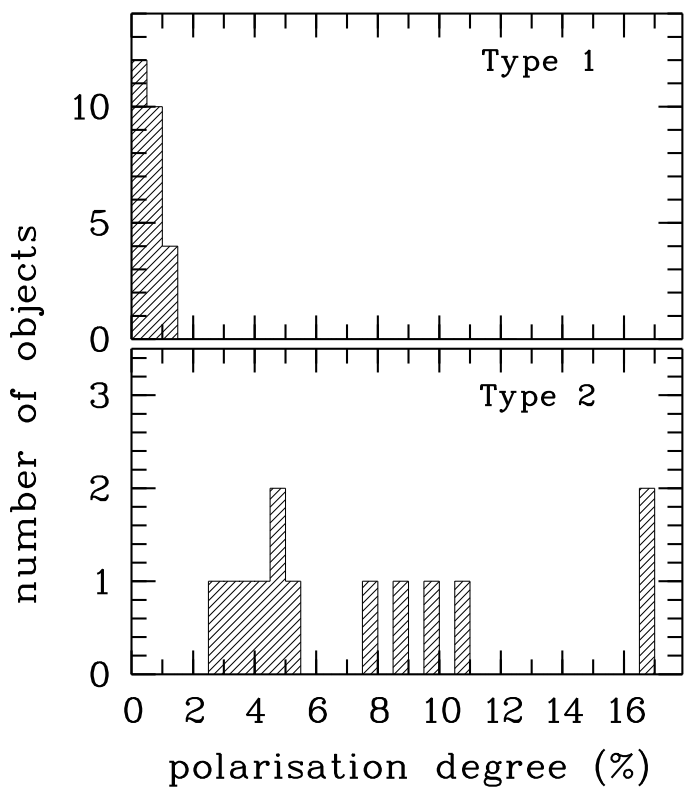}}
\caption{Distribution of the rest-frame UV-blue linear polarization in type~1 and~2 radio-quiet quasars with redshifts 0.2 $< z <$ 0.7. All type~1 quasars in that redshift range have $p < 2\%,$ while all type~2 quasars have $p > 2\%$.}
\label{fig:pola}
\end{figure}

The distributions of type~1 and~2 quasar polarizations are illustrated in Fig.~\ref{fig:pola}. A clear dichotomy is seen: all type~1 quasars have low polarizations, in agreement with previous studies  \citep[e.g.][]{1990Berriman}, while all type~2 quasars have high polarizations $p > 2\%$. It is important to note that the polarization measurements of type~1 quasars reported in Table~\ref{tab:pola} were obtained in the V filter (mean $\lambda \simeq$ 555 nm; \citealt{2005Sluse}) or in ``white light'' (mean $\lambda \simeq$ 590 nm; \citealt{1984Stockman, 1984Moore, 1990Berriman}), while the type~2 quasar polarizations obtained by \citet{2005Zakamska, 2006Zakamska} in the 2820-3710 \AA\ rest-frame wavelength range correspond to $g$-band observations (mean $\lambda \simeq$ 475 nm for an average redshift of 0.45). Since the polarization of type~1 quasars may slightly depend on wavelength as $p(\lambda) \propto \lambda^{\alpha}$ with $\alpha = -0.7 \pm 0.1$ \citep{1984Stockman}, we expect the $g$-band polarization of type~1 quasars to be at most 20\% higher than the values reported in Table~\ref{tab:pola}. This is negligible with respect to the observed type~1 / type~2 difference. A similar dichotomy is observed for lower luminosity Seyfert~1 and~2 AGNs \citep{2014Marin}, although it is not as sharp as for quasars. For lower luminosity objects with lower redshift, this might be due to higher contamination by the host galaxy and the fact that polarization is measured at longer rest-frame wavelengths\footnote{In addition, some Seyfert~1 AGNs show intermediate polarizations (polar-scattered Seyfert~1). Their high-luminosity counterparts might be broad absorption line quasars \citep{2004Smith}, which have been discarded from our comparison sample.}.

Following the models initially developped for Seyferts \citep[e.g.][]{2004Smith,2011Batcheldor,2014Marin}, the polarization of radio-quiet quasars can be interpreted by scattering off two regions: an equatorial ring located inside the dusty torus at the origin of ``parallel'' polarization, and a more extended polar-scattering region at the origin of ``perpendicular'' polarization \citep[e.g.][]{2005Zakamska,2008Borguet}. In type~1 quasars, seen at low inclinations (the angle between the system axis and the line of sight), the continuum source and both scattering regions are seen by the observer, resulting in a low polarization. In type~2 quasars, seen at high inclinations, the quasar core is hidden by the torus and only highly polarized polar-scattered light is seen. A change in polarization properties is thus expected from type~1 to type~2 depending on inclination, as is observed in Seyfert galaxies \citep{2004Smith,2014Marin}. A change in polarization is also expected if the torus properties are varying, as shown by \citet{2016Marin}. If the disappearance of broad emission lines in changing-look quasars is caused by torus dust clouds hiding the quasar core (i.e. the continuum source, the BEL region, and the equatorial scattering region), the high polarization typical of type~2 quasars is expected in quasars in which broad emission lines have disappeared.

\section{Polarization of J1011$+$5442}
\label{sec:obs}

SDSSJ101152.98$+$544206.4 (hereafter J1011$+$5442) is a changing-look quasar discovered by \citet{2016Runnoe}. SDSS photometric observations of J1011$+$5442 were obtained on February 13, 2002, and spectroscopic observations on January 13, 2003. These data led to the classification of  J1011$+$5442 as a broad-line radio-quiet type 1 quasar of redshift $z = 0.246$ and absolute magnitude $M_{i} = -22.87$ \citep{2005Schneider,2010Schneider,2011Shen}. J1011$+$5442 was re-observed spectroscopically on February 20, 2015, as part of the Time-Domain Spectroscopic Survey \citep{2015Morganson}, showing a complete disappearance of the strong broad H$\beta$ line \citep{2016Runnoe}. This spectroscopic change from type 1 to type 1.9 (faint broad H$\alpha$ is still detected) is accompanied by a systematic decline of the light curve from 2002 to 2015 \citep{2016Runnoe}. In particular, \citet{2016MacLeod} reported a brightness decrease from SDSS $g$ = 18.37 in February 2002 to Pan-STARRS1 $g$ = 19.88 in October 2011. A similar brightness change is found when comparing the GAIA G magnitude computed from the SDSS magnitudes in February 2002 using the colour transformation of \citet{2010Jordi}, that is, G = 18.03, to the magnitude reported in the GAIA Data Release~1 \citep{2016GAIADR1}, G = 19.91, which is the average of 40 observations between July 2014 and September 2015.

We obtained new linear polarization data of J1011$+$5442 on February 19, 2017, using the Intermediate dispersion Spectrograph and Imaging System (ISIS) mounted at the Cassegrain focus of the 4.2m William Herschel Telescope (WHT) at the Roque de los Muchachos Observatory. Observations were made through the blue arm and the Sloan Gunn $g$ filter (ING filter \#218; $\lambda_c$ = 4844 \AA , FWHM = 1280 \AA ), with ISIS in its imaging polarimetry mode. Four exposures of 15 minutes each with the half-wave plate at 8\degr, 30.5\degr, 53\degr, and 75.5\degr\ were secured. This sequence was repeated, but the last exposure was aborted due to increasing humidity. After bias subtraction and flat-fielding the frames, the Stokes parameters were measured using the procedures described in \citet{2005Sluse}. Polarized and unpolarized standard stars (HD283812, HD251204, and G191B2B from \citealt{1990Turnshek}) were observed to correct for the chromatic dependence of the half-wave plate zero-angle and to estimate the instrumental polarization, respectively. The linear polarization degree of J1011$+$5442 in the rest-frame 3200-4400~\AA\  wavelength range is found to be $p = 0.15 \%$ with an error $\sigma_p = 0.22 \%$. Since $\sigma_p > p$, the debiased polarization degree is equal to zero and the polarization position angle cannot be estimated  with reasonable accuracy. The instrumental polarization is measured equal to $p = 0.07 \pm 0.04 \%$. Based on the foreground Galactic extinction from \citet{2011Schlafly} given in the NASA/IPAC Extragalactic Database (NED), the interstellar polarization towards J1011$+$5442 is expected to be $p \leq 9\% \times E(B-V) \simeq 0.1\%$. The polarization of J1011$+$5442 is therefore compatible with a null intrinsic linear polarization.

To check the magnitude of J1011$+$5442 in February 2017, we used a faint star in the ISIS field of view that is located in one open slot of the mask: SDSSJ101153.55$+$544111.1, which  has a magnitude $g$ (SDSS) = 22.1$\pm$0.1. By stacking all frames obtained in the polarimetric mode, we measured $F_{\text{quasar}} / F_{\text{star}} = 10.5 \pm 1.0$ in the WHT $g$ filter. This translates into $g$ $\simeq$ 19.6 $\pm$ 0.2 for J1011$+$5442. Although this is a rough estimate, it indicates that J1011$+$5442 was still in its faint state in February 2017.

\section{Discussion and conclusion}
\label{sec:end}

Although the quasar J1011$+$5442 is in a type~2 dim state, its polarization appears to be very low, typical of unobscured type~1 quasars. This suggests that the observed change of look is not due to a change of obscuration in the torus hiding the continuum source, the broad line region, and the equatorial scattering region. The null polarization also suggests that the quasar is seen at low inclination so that the scattering regions appear essentially symmetric, that is, far from lines of sight crossing the dusty torus. Our results thus support the idea that the vanishing of the broad emission lines is due to an intrinsic dimming of the ionizing continuum source most likely caused by a rapid decrease of the rate of accretion onto the supermassive black hole, in agreement with the conclusions reached independently by \citet{2015LaMassa}, \citet{2016Runnoe}, and \citet{2016MacLeod}. In this scenario, the brightness decrease of the continuum source would be accompanied by a similar decrease of the scattered flux, with no significant change of the polarization degree. A time delay is nevertheless expected between the variations of the direct and scattered light, depending on the sizes of the equatorial and polar scattering regions.  The size of the equatorial scattering region being comparable to that of the broad line region, the time delay is about one month in  J1011$+$5442 \citep{2016Runnoe}, which is short with respect to the timescale of the change of look (years). The size of the polar scattering region, on the other hand, is much larger. While the time delay between the direct and polar-scattered light remains small at low inclination, it could then be larger at intermediate inclinations, possibly generating complex polarization variations. This is the basis of polarization reverberation mapping \citep[e.g.][]{2012Gaskell}.

Polarimetry of changing-look quasars thus appears useful not only to distinguish the mechanisms at the origin of quasar type
changes, but also to constrain the properties of the scattering regions. This implicitly assumes that changing-look quasars, and J1011$+$5442 in particular, do possess the scattering regions found in other radio-quiet quasars. A monitoring of the linear polarization of a significant sample of changing-look quasars is thus needed.

\begin{acknowledgements}
DH and BAG thank Marie Hrudkova for her help during the observations. CRA acknowledges the Ram\'on y Cajal Program of the Spanish Ministry of Economy and Competitiveness through project RYC-2014-15779 and the Spanish Plan Nacional de Astronom\' ia y Astrofis\' ica under grant AYA2016-76682-C3-2-P. This research has made use of the NASA/IPAC Extragalactic Database (NED) which is operated by the Jet Propulsion Laboratory, California Institute of Technology, under contract with the National Aeronautics and Space Administration. This work has made use of data from the European Space Agency (ESA) mission {\it Gaia} (\url{https://www.cosmos.esa.int/gaia}), processed by the {\it Gaia} Data Processing and Analysis Consortium (DPAC, \url{https://www.cosmos.esa.int/web/gaia/dpac/consortium}). Funding for the DPAC has been provided by national institutions, in particular the institutions participating in the {\it Gaia} Multilateral Agreement.
\end{acknowledgements}

\bibliographystyle{aa}
\bibliography{references}

\end{document}